\documentclass[aip,jcp,reprint,amsmath,amssymb,floatfix,citeautoscript]{revtex4-1}
\usepackage{cancel}
\usepackage{amsmath}
\usepackage{graphicx}
\usepackage{amssymb}
\usepackage{amsthm}
\usepackage{bm}
\usepackage{dcolumn}

\usepackage{ragged2e}
\usepackage[labelsep=period,format=plain,justification=justified,font=footnotesize]{caption}
\DeclareCaptionJustification{myjust}{\justifying}
\usepackage{txfonts}

\usepackage{color}
\usepackage[usenames,dvipsnames]{xcolor}
\definecolor{myblue}{rgb}{0,0,1}
\usepackage[breaklinks=true,colorlinks=true,linkcolor=myblue,urlcolor=myblue,citecolor=myblue]{hyperref}

\let\vr\undefined
\newcommand{\vr}{{\bm{r}}}

\newcommand{\vk}{{\bm{k}}}

\newcommand{\vq}{{\bm{q}}}
\newcommand{\vR}{{\bm{R}}}
\newcommand{\vG}{{\bm{G}}}

\newcommand{\vT}{{\bm{T}}}

\newcommand{\ii}{i\vk_i}
\newcommand{\jj}{j\vk_j}
\newcommand{\kk}{k\vk_k}
\let\ll\undefined
\newcommand{\ll}{l\vk_l}
\let\aa\undefined
\newcommand{\aa}{a\vk_a}
\newcommand{\bb}{b\vk_b}
\newcommand{\cc}{c\vk_c}
\newcommand{\dd}{d\vk_d}
\newcommand{\pp}{p\vk_p}
\newcommand{\qq}{q\vk_q}
\newcommand{\rr}{r\vk_r}
\let\ss\undefined
\newcommand{\ss}{s\vk_s}
\newcommand{\rip}[3]{{r_{#1#2}^{\hphantom{#1}#3}}}
\newcommand{\rea}[3]{{r_{\hphantom{#2}#1}^{#2#3}}}

\begin{document}

\title{Gaussian-based coupled-cluster theory for the ground state and band structure of
solids}

\author{James McClain}
\affiliation{Department of Chemistry, Princeton University, 
Princeton, New Jersey 08544, USA}

\author{Qiming Sun}
\affiliation{Division of Chemistry and Chemical Engineering, 
California Institute of Technology, Pasadena, California 91125, USA}

\author{Garnet Kin-Lic Chan}
\email{gkc1000@gmail.com}
\affiliation{Division of Chemistry and Chemical Engineering, 
California Institute of Technology, Pasadena, California 91125, USA}

\author{Timothy C. Berkelbach}
\email{berkelbach@uchicago.edu}
\affiliation{Department of Chemistry and James Franck Institute,
University of Chicago, Chicago, Illinois 60637, USA}

\begin{abstract}

We present the results of Gaussian-based ground-state and excited-state equation-of-motion
coupled-cluster theory with single and double excitations for three-dimensional solids.
We focus on diamond and silicon, which are paradigmatic covalent semiconductors.  In
addition to ground-state properties (the lattice constant, bulk modulus, and cohesive
energy), we compute the quasiparticle band structure and band gap.  We sample the
Brillouin zone with up to 64 $k$-points using norm-conserving pseudopotentials and
polarized double- and triple-zeta basis sets, leading to canonical coupled-cluster
calculations with as many as 256 electrons in 2,176 orbitals.

\end{abstract}

\maketitle

\section{Introduction}

The electronic ground-state and excited-state structure of three-dimensional solids and
low-dimensional nanomaterials is an ongoing challenge in computational materials science.
A theoretical program based on density functional theory (DFT), frequently combined with
time-dependent many-body perturbation theory, has become a reliable and accurate standard
approach.  In particular, ground-state properties of weakly correlated materials are
well-described by DFT, the principal charged excited states leading to band structure can
be treated with hybrid~\cite{Heyd2003,Clark2010,Crowley2016} or semilocal~\cite{Tran2009}
functionals, as well as the many-body $GW$
approximation~\cite{Hybertsen1985,Hybertsen1986}, and neutral excited states can be
described with time-dependent DFT~\cite{deBoeij2001,Reining2002} or the Bethe-Salpeter
equation~\cite{Rohlfing2000}.  However, errors in this approach can be difficult to
assess, due to (for example) the underlying DFT exchange-correlation functional or
self-consistency in the $GW$ approximation.  Even in the formal many-body frameworks,
systematic improvements (vertex corrections) are only recently being pursued, with mixed
success~\cite{Shishkin2007PRL,Romaniello2009,Gruneis2014,Pavlyukh2016}.  

In contrast, wavefunction-based approaches originating in the quantum chemistry community
are now starting to be applied to the properties of condensed-phase materials.  Atomistic
one- and two-dimensional systems have been studied by configuration interaction with
single excitations (CIS), time-dependent Hartree-Fock (HF), and time-dependent
DFT~\cite{Hirata1999}, as well as second-order M\o ller-Plesset perturbation theory
(MP2)~\cite{Sun1996,Ayala2001,Hirata2001,Hirata2004} and coupled-cluster (CC) theory with
single and double excitations (CCSD)~\cite{Sun1996,Hirata2001,Hirata2004}.  For
three-dimensional solids, MP2 has been implemented with local correlation in the CRYSCOR
program~\cite{Pisani2003,Pisani2005}, in the plane-wave based VASP package with full
Brillouin zone sampling\cite{Marsman2009,Gruneis2010,Gruneis2011}, and in the mixed
Gaussian and plane-wave CP2K package at the $\Gamma$-point~\cite{DelBen2012}.  Most
recently, ground-state CCSD, CCSD with perturbative triple excitations (CCSD(T)), and full
CI quantum Monte Carlo for periodic solids have been
reported~\cite{Gruneis2011,Booth2013}.  We also mention the parallel development of
non-periodic cluster-based approaches for solids, namely incremental
methods~\cite{Stoll1991,Stoll2005,Paulus2006,Nolan2009} and embedding
techniques~\cite{Libisch2014,Kotliar2006,Sun2016}.

Recently, we performed the first excited-state CC treatment of a three-dimensional
condensed-phase system, at the equation-of-motion (EOM)-CCSD and EOM-CCSDT level, focusing
on the uniform electron gas at the density of metallic sodium~\cite{McClain2016}.  For
small unit cells in a finite single-particle basis, the EOM-CCSDT one-particle spectral
function was in nearly perfect agreement with numerically exact (time-dependent density
matrix renormalization group) results, and more accurate than any tested flavor of the
$GW$ approximation, while the EOM-CCSD one-particle spectral function quantitatively
improved on the standard $GW$ approximation.  Importantly, unlike the $GW$ approximation
(in practice, $G_0W_0$), the EOM-CC spectra are invariant to the starting choice of
single-particle energies.  At larger system sizes approaching the thermodynamic limit, it
was not possible to carry out full EOM-CCSDT calculations, but it was still possible to
compute the EOM-CCSD spectrum.  Similarly to the $GW$-plus-cumulant
approximation~\cite{Hedin1980,Aryasetiawan1996}, the EOM-CCSD spectrum removed the
fictitious plasmaron pole of the $GW$ spectrum~\cite{Hedin1967,Lundqvist1967},
demonstrating that it correctly treats the coupling of charged excitations to bosonic
plasmon excitations in metals.

Here, we continue this program and present the first results of CC excited states for
atomistic solids.  The layout of this paper is as follows. In Sec.~\ref{sec:hf}, we
describe the Gaussian basis sets, integral evaluation, and HF calculations for periodic
systems, including issues associated with divergent terms and finite-size effects.  In
Sec.~\ref{sec:cc}, we describe the ground-state CCSD and excited-state EOM-CCSD formalisms
for periodic systems, as well as the connection to quasiparticle band structure.  In
Sec.~\ref{sec:finite} we discuss the convergence and finite-size effects of our HF and
correlated calculations.  We present ground-state HF, MP2, and CCSD results in
Sec.~\ref{sec:results_gs} for diamond and silicon.  This section also includes the
technical details of our calculations, a discussion of convergence, and comparison to
existing MP2 and CCSD calculations on solids~\cite{Gruneis2010,Booth2013}. In
Sec.~\ref{sec:results_eom}, we present EOM-CCSD band structures and a convergence study of
the indirect band gap.  We conclude in Sec.~\ref{sec:conc}.

\section{Gaussian-based periodic integrals and Hartree-Fock theory}
\label{sec:hf}

We use an underlying single-particle basis of crystalline Gaussian-based atomic orbitals
(AOs). These are translational-symmetry-adapted linear combinations of Gaussian AOs of the
form
\begin{equation}
\label{eq:crystalline_ao}
\phi_{\mu\vk}(\vr) = \sum_{\vT} e^{i\vk\cdot\vT} \tilde{\phi}_{\mu}(\vr-\vT) 
    \equiv e^{i\vk\cdot\vr} u_{\mu\vk}(\vr)
\end{equation}
where $\vT$ is a latice translation vector and $\vk$ is a crystal momentum vector in the
first Brillouin zone. In this work, we sample $\vk$ from a uniform (but not necessarily
Monkhorst-Pack) grid. The second equality above expresses Bloch's theorem, where the Bloch
function $u_{\mu\vk}(\vr)$ is fully periodic with respect to all lattice translations.
Therefore, we can exactly expand the crystalline AOs in a set of auxiliary plane-waves
\begin{subequations}
\label{eq:ao_pw}
\begin{align}
    \phi_{\mu\vk}(\vr) &= \sum_\vG \phi_{\mu\vk}(\vG)e^{i(\vk+\vG)\cdot\vr}, \\
    \phi_{\mu\vk}(\vG) &= \frac{1}{\Omega}\int_\Omega d\vr \phi_{\mu\vk}(\vr)e^{-i(\vk+\vG)\cdot\vr},
\end{align}
\end{subequations}
where $\vG$ is a reciprocal lattice vector and $\Omega$ is the unit cell volume.

Given this basis choice, there are several different ways to evaluate the corresponding
overlap, kinetic energy, electron-nuclear attraction, and Coulomb integrals arising in
quantum chemistry.  We will present a careful comparison of different choices in a later
publication, and here only detail the grid-based scheme that we have used for the results
in this work, which closely resembles the Gaussian and plane-wave scheme used in the CP2K
package~\cite{VandeVondele2005,CP2K}.  We use a dual real-space and reciprocal-space
representation of the crystalline AOs.  In real space, we represent $\phi_{\mu\vk}(\vr)$
on a uniform real-space grid in the unit cell.  In reciprocal space, we obtain
$\phi_{\mu\vk}(\vG)$ by a fast Fourier transform (FFT) leading to a representation on a
uniform grid of points corresponding to the reciprocal lattice vectors of our unit cell.
The real-space grid density roughly corresponds to a kinetic energy cutoff in the
plane-wave representation.  To ensure that relatively low cutoffs can be used, we replace
the core electrons with separable norm-conserving GTH (HGH)
pseudopotentials~\cite{Goedecker1996,Hartwigsen1998}, which removes the sharp nuclear
densities. We use the Gaussian basis sets that are designed for use with these
pseudopotentials in solid-state calculations~\cite{VandeVondele2005}. 

The one-electron overlap, kinetic energy, and local (L) pseudopotential integrals are
evaluated by numerical integration on the real-space grid according to
\begin{align}
S_{\mu\nu}(\vk) &= \int_\Omega d\vr \phi^*_{\mu\vk}(\vr) \phi_{\nu\vk}(\vr), \\
T_{\mu\nu}(\vk) &= -\frac{1}{2}\int_\Omega d\vr \phi^*_{\mu\vk}(\vr) \nabla_\vr^2 \phi_{\nu\vk}(\vr), \\
V^{\mathrm{L}}_{\mu\nu}(\vk) &= \int_\Omega d\vr \phi^*_{\mu\vk}(\vr) v^{\mathrm{L}}(\vr) \phi_{\nu\vk}(\vr);
\end{align}
where we note that these integrals and other quantities throughout the paper are defined
per unit cell.  The total, local contribution from the ion pseudopotentials
$v^{\mathrm{L}}(\vr)$ is given by,
\begin{equation}
v^{\mathrm{L}}(\vr) = \sum_{\vG \neq 0} \sum_I e^{i\vG\cdot(\vr-\vR_I)} v^{\mathrm{L}}_I(\vG),
\end{equation}
where $I$ indexes the ions at position $\vR_I$ in the unit cell.  The local part of the
GTH pseudopotential is finite at $r=0$ but decays as $-Z_I/r$ leading to a divergent $G=0$
component, which is treated separately (see below).  The nonlocal (NL) pseudopotential
part is evaluated on the reciprocal-space grid
\begin{equation}
V^{\mathrm{NL}}_{\mu\nu}(\vk) = \Omega \sum_{\vG,\vG^\prime}
    \phi_{\mu\vk}^*(\vG)v^{\mathrm{NL}}(\vk+\vG,\vk+\vG^\prime) \phi_{\nu\vk}(\vG^\prime)
\end{equation}
with
\begin{equation}
v^{\mathrm{NL}}(\vk+\vG,\vk+\vG^\prime) = \sum_I 
    e^{-i(\vG-\vG^\prime)\cdot\vR_I} v^{\mathrm{NL}}_I(\vk+\vG,\vk+\vG^\prime)
\end{equation}
and
\begin{equation}
\begin{split}
v^{\mathrm{NL}}_I(\vk+\vG,\vk+\vG^\prime) &= \frac{1}{\Omega}
    \int d\vr \int d\vr^\prime e^{-i(\vk+\vG)\cdot\vr} \\
&\hspace{4em} \times v^{\mathrm{NL}}_I(\vr,\vr^\prime) e^{i(\vk+\vG^\prime)\cdot\vr^\prime}.
\end{split}
\end{equation}
In the above, the ion-specific local and nonlocal parts of the pseudopotential have
analytic definitions of the GTH form~\cite{Goedecker1996,Hartwigsen1998}. In particular
the nonlocal part is separable and expressible as a sum of products of functions of
$\vk+\vG$ and functions of $\vk+\vG^\prime$.

The Hartree and exchange matrices are evaluated in real space,
\begin{align}
J_{\mu\nu}(\vk) &= \int_\Omega d\vr \phi^*_{\mu\vk}(\vr) v_{\mathrm{H}}(\vr) \phi_{\nu\vk}(\vr), \\
K_{\mu\nu}(\vk) &= \int_\Omega d\vr \int d\vr^\prime 
    \phi^*_{\mu\vk}(\vr) \frac{\rho(\vr,\vr^\prime)}{|\vr-\vr^\prime|} \phi_{\nu\vk}(\vr^\prime),
\end{align}
in terms of the Hartree potential
\begin{equation}
v_{\mathrm{H}}(\vr) 
    = \frac{4\pi}{\Omega} \sum_{\vG\neq 0} \frac{\rho(\vG)}{G^2} e^{i\vG\cdot\vr} 
\end{equation}
and the density matrix $\rho(\vr,\vr^\prime)$. In general, the density matrix and density
$\rho(\vr) = \rho(\vr,\vr)$ can be obtained from a Brillouin zone sampling
\begin{equation}
\label{eq:dm}
\rho(\vr,\vr^\prime) = \frac{1}{N_k} \sum_{\vk} \sum_{\lambda\sigma} P_{\lambda\sigma}(\vk) \phi_{\lambda\vk}(\vr) \phi_{\sigma\vk}^*(\vr^\prime), 
\end{equation}
where $N_k$ is the number of $k$-points sampled in the Brillouin zone.  Assuming a
closed-shell reference of molecular orbitals, 
$\psi_{p\vk} = \sum_\mu C_{\mu p}(\vk) \phi_{\mu\vk}$, the 
mean-field density matrix is given by
\begin{equation}
P_{\lambda\sigma}(\vk) = 2 \sum_i^\mathrm{occ} C_{\lambda i}(\vk) C_{\sigma i}^*(\vk).
\end{equation}
This leads to the real-space form of the exchange matrix
\begin{equation}
\label{eq:K}
K_{\mu\nu}(\vk) = \frac{1}{N_k} \sum_{\vk^\prime} \sum_{\lambda\sigma} P_{\lambda\sigma}(\vk^\prime) 
    \int d\vr \phi^*_{\mu\vk}(\vr) v^{\mathrm{X}}_{\nu\vk,\sigma\vk^\prime}(\vr) \phi_{\lambda\vk^\prime}(\vr) 
\end{equation}
where
\begin{equation}
\begin{split}
\label{eq:vX}
v^{\mathrm{X}}_{\sigma\vk^\prime,\nu\vk}(\vr) &= \int d\vr^\prime 
    \frac{\phi^*_{\sigma\vk^\prime}(\vr^\prime) \phi_{\nu\vk}(\vr^\prime)}{|\vr-\vr^\prime|}
    \equiv \int d\vr^\prime 
        \frac{\rho_{\sigma\vk^\prime,\nu\vk}(\vr^\prime)}{|\vr-\vr^\prime|} \\
&= \frac{4\pi}{\Omega} \sum_{\vG}^\prime \frac{\rho_{\sigma\vk^\prime,\nu\vk}(\vG)}{|\vk-\vk^\prime+\vG|^2}
    e^{i(\vk-\vk^\prime+\vG)\cdot \vr},
\end{split}
\end{equation}
and the $G=0$ term is excluded when $\vk=\vk^\prime$.
Like the individual crystalline AOs, the AO pair densities have a plane-wave resolution,
\begin{subequations}
\label{eq:density_pw}
\begin{align}
\begin{split}
\rho_{\sigma\vk^\prime,\nu\vk}(\vr) &= \phi^*_{\sigma\vk^\prime}(\vr)\phi_{\nu\vk}(\vr) \\
&= \sum_\vG \rho_{\sigma\vk^\prime,\nu\vk}(\vG) e^{i(\vk-\vk^\prime+\vG)\cdot \vr},
\end{split}\\
\begin{split}
\rho_{\sigma\vk^\prime,\nu\vk}(\vG) &= \frac{1}{\Omega} \int_{\Omega}
    \rho_{\sigma\vk^\prime,\nu\vk}(\vr) e^{-i(\vk-\vk^\prime+\vG)\cdot \vr}.
\end{split}
\end{align}
\end{subequations}

As shown above, the local part of the pseudopotential and Hartree potential are separately
divergent at $G=0$, however their sum is not. For each atom, the sum
is given by
\begin{equation}
\alpha_I = \int d\vr \left(v^{\mathrm{L}}_I(\vr) + \frac{Z_I e^2}{r}\right),
\end{equation}
where $\alpha_I$ is a finite, ion-specific parameter of the pseudopotential.  
This leads to a additional matrix element
\begin{equation}
V_{\mu\nu}^{\mathrm{L}+J}(\vk) = \frac{S_{\mu\nu}(\vk)}{\Omega}\sum_I \alpha_I,
\end{equation}
resulting simply in a uniform shift of the orbital energies and a constant in the total
energy.  Note that in the absence of pseudopotentials, the divergent contributions from
the Hartree and electron-nuclear interaction cancel exactly by charge neutrality. 

Once the above integrals are defined, then it is straightforward to carry out a HF
calculation using these integrals. The only difference compared to a standard molecular HF
calculation is that the integrals and orbitals are complex.  The molecular orbitals at
each $k$-point $\psi_{p\vk}$ are obtained from the HF equation
\begin{equation}
\mathbf{F}(\vk) \mathbf{C}(\vk) = \mathbf{\varepsilon}(\vk) \mathbf{S}(\vk) \mathbf{C}(\vk)
\end{equation}
where
\begin{equation}
\mathbf{F}(\vk) = \mathbf{T}(\vk) 
    + \mathbf{V}^{\mathrm{PP}}(\vk)
    + \mathbf{J}(\vk) -\tfrac{1}{2} \mathbf{K}(\vk)
    + \mathbf{V}^{\mathrm{L}+J}(\vk)
\end{equation}
and $\mathbf{V}^{\mathrm{PP}}(\vk) = \mathbf{V}^{\mathrm{L}}(\vk) 
+ \mathbf{V}^{\mathrm{NL}}(\vk)$.
Assuming $n_{\mathrm{el}}$ electrons per unit cell, the HF determinant includes the $N_k
n_{\mathrm{el}}$ orbitals with the lowest eigenvalues out of all $k$-points.  The total
energy is the usual HF one plus the nuclear-nuclear repulsion, which is computed using the
Ewald expression 
\begin{equation}
\begin{split}
E_{\mathrm{NN}} &= \frac{1}{2} \sum_{IJ\vT}^\prime \frac{Z_I Z_J}{|\vR_I - \vR_J - \vT|} 
    \mathrm{erfc}\left(\eta |\vR_I - \vR_J - \vT|\right) \\
&\hspace{1em} + \frac{1}{2} \frac{4\pi}{\Omega} \sum_{\vG\neq 0} 
    \left| \sum_I Z_I e^{i\vG\cdot\vR_I} \right|^2 \frac{e^{-G^2/4\eta^2}}{G^2} \\
&\hspace{1em} - \frac{\eta}{\sqrt{\pi}} \sum_I Z_I^2
    - \frac{\pi}{2\Omega\eta^2} \left(\sum_I Z_I\right)^2,
\end{split}
\end{equation}
where the primed summation neglects the self-interaction terms with $I=J$ when $\vT = 0$
and the range of lattice summations is chosen together with $\eta$ to facilitate rapid
convergence. 

For a subsequent correlation treatment, we need to define the molecular orbital integrals.
The one-electron integrals can be obtained straightforwardly by changing basis with 
$\mathbf{C}(\vk)$.  
Two-electron integrals are defined (again, per unit cell) as
\begin{equation}
\begin{split}
&\langle \pp, \rr | \qq, \ss\rangle = (\pp, \qq | \rr, \ss) \\
&\hspace{0.5em} = \int_\Omega d\vr_1 \int d\vr_2 
    \psi_{\pp}^*(\vr_1) \psi_{\qq}(\vr_1) v(r_{12})
        \psi^*_{\rr}(\vr_2) \psi_{\ss}(\vr_2). 
\end{split}
\end{equation}
In the presence of translational invariance, these two-electron integrals
must conserve crystal momentum, i.e.~$\vk_p + \vk_r - \vk_q - \vk_s = \vG$,
where $\vG$ is a reciprocal lattice vector.  We evaluate these integrals by representing
the molecular orbital pair density in terms of plane-waves, leading to
\begin{equation}
\label{eq:eris}
\begin{split}
& (\pp, \qq | \rr, \ss) = \Omega^2 \sum^\prime_\vG
    \Big[ \rho_{\pp,\qq}(\vG) \\
&\hspace{6em} \times v(\vk_q-\vk_p+\vG) \rho_{\rr,\ss}(\vG_{prqs}-\vG) \Big].
\end{split}
\end{equation}
where $G_{prqs} = \vk_p + \vk_r - \vk_q - \vk_s$ is a reciprocal lattice vector,
$v(\vG) = 4\pi/\Omega G^2$, and the $G=0$ singularity is removed when $\vk_p=\vk_q$. 

Once these integrals are obtained, we can define the second-quantized Hamiltonian suitable
for a correlated electronic structure treatment.  For closed-shell single-reference
correlation techniques, it is convenient to express the Hamiltonian in normal-ordered form
\begin{equation}
\begin{split}
H &= \langle \Phi | H | \Phi\rangle
    + \sum_{pq}\sum_{\vk} F_{pq}(\vk) \left\{E_{q\vk}^{p\vk}\right\} \\
&\hspace{1em} + \frac{1}{2} \sum_{pqrs} \sum_{\vk_p\vk_q\vk_r\vk_s}^\prime
    (\pp \qq | \rr \ss) \left\{E_{\qq}^{\pp} E_{\ss}^{\rr}\right\}
\end{split}
\end{equation}
where $E_{\qq}^{\pp} = \sum_{\sigma} a_{\pp\sigma}^\dagger a_{\qq\sigma}$ is a spin-summed
excitation operator, $\{ \dots \}$ denotes normal ordering with respect to the reference
$\Phi$ (the Fermi vacuum), and as usual the primed summation enforces crystal momentum
conservation.  Naturally, in the case of canonical HF, the Fock matrix is diagonal,
$F_{pq}(\vk) = \varepsilon_p(\vk)\delta_{pq}$.

\section{Periodic EOM coupled-cluster}
\label{sec:cc}

Given a closed-shell reference determinant $|\Phi\rangle$, the ground-state CCSD
wavefunction is $|\Psi_0\rangle = e^{T} |\Phi\rangle$ where $T=T_1+T_2$, and $T_1$ and
$T_2$ are single- and double-excitation
operators~\cite{Cizek1966,Scuseria1988,ShavittBartlettBook,Bartlett2007}
\begin{subequations}
\label{eq:t1t2}
\begin{align}
  T_1 &= \sum_{ai} \sum_{\vk_a \vk_i}^\prime t_{\ii}^{\aa} E_{\ii}^{\aa}, \\
  T_2 &= \frac{1}{2} \sum_{abij} \sum_{\vk_a \vk_b \vk_i \vk_j}^\prime 
        t_{\ii \jj}^{\aa \bb} E_{\ii}^{\aa} E_{\jj}^{\bb}. 
\end{align}
\end{subequations}
In this work, we use a HF reference determinant, although the equations below and in the
Appendix apply to any determinant.  In Eqs.~(\ref{eq:t1t2}) and throughout, the indices
$i,j,k,l$ denote occupied orbitals and $a,b,c,d$ denote virtual orbitals.  Because the
crystal Hamiltonian has translational symmetry, the excitation operators must conserve
crystal momentum, i.e.  $\sum_a \vk_a - \sum_i \vk_i = \bm{G}$ where $\vk_a$ and $\vk_i$
are the crystal momenta of particle and hole orbitals and $\bm{G}$ is a reciprocal lattice
vector.  This requirement is indicated by the primed summation in Eq.~(\ref{eq:t1t2}) and
we emphasize that each primed summation indicates that one of the listed momenta is fixed
and need not be summed.  Introducing the non-Hermitian CC effective Hamiltonian $\bar{H}
\equiv e^{-T}He^{T}$, the CCSD ground-state energy and excitation amplitudes are
determined by
\begin{align}
E_0 &= \langle \Phi_0 | \bar{H} |\Phi_0\rangle, \\
0   &= \langle \Phi_{\ii\alpha}^{\aa\alpha} | \bar{H} |\Phi_0\rangle, \tag{\addtocounter{equation}{1}\theequation a} \\
0   &= \langle \Phi_{\ii\alpha,\jj\beta}^{\aa\alpha,\bb\beta} | \bar{H} |\Phi_0\rangle, \tag{\theequation b}
\end{align}
where $\Phi_{\ii\alpha}^{\aa\alpha}$ and $\Phi_{\ii\alpha,\jj\beta}^{\aa\alpha,\bb\beta}$
are Slater determinants with one and two electron-hole pairs. The explicit forms of the
energy and amplitude equations with translational symmetry are given in
Ref.~\onlinecite{Hirata2004}.  Importantly, the computational cost of CCSD for periodic
systems scales as $n_{\mathrm{occ}}^2 n_{\mathrm{vir}}^4 N_k^4$, where $n_{\mathrm{occ}}$
and $n_{\mathrm{vir}}$ denote the number of occupied and virtual orbitals per unit cell;
this is a factor of $N_k^2$ less than the equivalent calculation that neglects momentum
conservation.

Coupled-cluster excited states and energies are determined through the equation-of-motion
(EOM) formalism~\cite{Monkhorst1977,Stanton1993,Krylov2008}, which amounts to
diagonalizing the effective Hamiltonian $\bar{H}$ in an appropriate space of excitations.
For a CCSD ground state, we calculate ionization potentials (IPs) via diagonalization in
the space of 1-hole ($1h$) and 2-hole, 1-particle ($2h1p$) states, and we calculate
electron affinities (EAs) via diagonalization in the space of 1-particle ($1p$) and
2-particle, 1-hole ($2p1h$) states:
$|\Psi_{n,\vk}^{N\pm 1}\rangle = R^\pm(n,\vk) |\Psi_0\rangle 
= [R^\pm_1(n,\vk)+R^\pm_2(n,\vk)]|\Psi_0\rangle$
where $R^-_1$ ($R^+_1$) creates $1h$ ($1p$) excitations and $R^-_2$ ($R^+_2$) creates
$2h1p$ ($2p1h$) excitations in the IP-EOM (EA-EOM) framework.  Explicitly,
\begin{align}
R^-_{\alpha}(n,\vk) &= \sum_i r_{i\vk} a_{i\vk\alpha}
    + \sum_{bij} \sum_{\vk_b \vk_i \vk_j}^\prime 
        \rip{\ii}{\jj}{\bb} E_{\jj}^{\bb} a_{\ii\alpha} \\
R^+_{\alpha}(n,\vk) &= \sum_a r^{a\vk} a^\dagger_{a\vk\alpha}
    + \sum_{abj} \sum_{\vk_a \vk_b \vk_j}^\prime 
        \rea{\jj}{\aa}{\bb} a_{\aa\alpha}^\dagger E_{\jj}^{\bb}
\end{align}
where again the primed summations are restricted to enforce momentum conservation.  The
$r$-amplitudes satisfy the usual IP/EA EOM equations, modified to include crystal momenta;
explicit expressions are given in the Appendix.  Again we emphasize that \textit{each}
primed summation leads to one crystal momentum which can be fixed by momentum
conservation; for example, in the final term of the $R_2$ amplitude equations
(\ref{eq:r2_ip}) and (\ref{eq:r2_ea}), only two of the four crystal momenta need to be
explicitly summed.
Note that although ground-state CCSD scales as $N^6$, the subsequent EOM-CCSD calculations
have a reduced scaling of $N^5$.  In particular, in the presence of periodicity, the
EOM-CCSD scaling for all charged excitations \textit{at a given $k$-point} is
$n_{\mathrm{occ}}^3 n_{\mathrm{vir}}^2 N_k^3$ for IP-EOM-CCSD and $n_{\mathrm{occ}}
n_{\mathrm{vir}}^4 N_k^3$ for EA-EOM-CCSD (if the charged excitations at all sampled
$k$-points are desired, then $N_k$ such calculations can be performed independently).
This specific scaling with the number of occupied and virtual orbitals usually makes the
calculation of conduction (virtual) bands significantly more expensive than that of
valence (occupied) bands.

In this work we focus on the quasiparticle excitations. These are the many-body states
that have the largest overlap with the mean-field single-particle excitations, i.e.~the
largest $|r_{i,\vk}|^2$ and $|r^{a,\vk}|^2$ (a proxy for the EOM Green's function pole
strength). This observation leads to an efficient targeted Davidson diagonalization
procedure based on the wavefunction character, rather than the
energy~\cite{Butscher1976,Zuev2015}.  At each $k$-point sampled in the Brillouin zone, we
typically target the three lowest-lying IP and EA excitations with such single-particle
character.  Away from the band edge, the quasiparticle picture breaks down: the
excitations develop an effective linewidth and nontrivial satellite structure through a
growing contribution of multiple determinants, which also leads to slower convergence of
the Davidson procedure. In this regime, it is more efficient to directly construct the
momentum- and frequency-dependent, one-particle Green's function as done in our previous
work on the uniform electron gas~\cite{McClain2016}.

\section{Convergence and finite-size effects}
\label{sec:finite}

It is worth briefly discussing the requirements for converging the correlated electronic
structure of a periodic system.  As in a molecular electronic structure calculation, we
must converge with respect to the single-particle basis set, i.e.~the set of Gaussian
functions $\tilde{\phi}_\mu(\vr)$ in Eq.~(\ref{eq:crystalline_ao}), as well as with
respect to the correlation level.  Convergence in these two regards can be expected to be
similar to that in a molecular system.  
However, in addition, we must converge with respect to the Brillouin zone sampling,
i.e.~the set of crystal momentum vectors $\vk$ appearing in Eq.~(\ref{eq:crystalline_ao}).
(This may formally be thought of as part of the basis set convergence for the infinite
system).  There is substantial experience with this convergence behavior at the mean-field
level (HF and DFT), but much less so at the correlated
level~\cite{Marsman2009,Gruneis2010,Gruneis2011,Liao2016}. However, it is important to
note that convergence with respect to Brillouin zone sampling can be quite slow. 

In general, if all quantities are smooth in the Brillouin zone, then approximating
integrals by finite Brillouin zone sampling has exponentially small error in the grid
spacing, i.e.~$\exp(-aN_k^{-1/3})$.  However, omitting the $G=0$ contribution in the
definition of Coulomb-based integrals is an additional source of finite-size error.  
For example, the exact HF exchange energy per unit cell is given by
\begin{equation}
E_\mathrm{X} = - \sum_{ij} \int_{\mathrm{BZ}} \frac{\Omega d\vk_i}{(2\pi)^3}
    \int_{\mathrm{BZ}} \frac{\Omega d\vk_j}{(2\pi)^3} (\ii \jj | \jj \ii).
\end{equation}
In this limit of infinite $k$-point sampling, the volume element associated with Brillouin
zone integration cancels the divergent Coulomb term, i.e.~it is an integrable divergence.
However, using the molecular orbital integrals from Eq.~(\ref{eq:eris}) and a finite
Brillouin zone sampling leads to the approximate exchange energy
\begin{equation}
\begin{split}
E_\mathrm{X} = -\frac{\Omega}{N_k^2} \sum_{ij} \sum_{\vk_i \vk_j \vG}^\prime 
    \left|\rho_{\ii,\jj}(\vG)\right|^2 \frac{4\pi}{|\vG-\vq_{ij}|^2},
\end{split}
\end{equation}
where the primed summation excludes $G=0$ when $\vq_{ij} = \vk_i-\vk_j = 0$.  This
neglected term represents an $O(1)$ integrand coming from the nonzero charge density
$\rho_{i\vk,i\vk}(\vG=0)$, leading to an integration error of $O(N_k^{-1/3})$.  In these
cases related to HF exchange, various corrections exist that aim to accelerate the
Brillouin zone convergence, including auxiliary function techniques~\cite{Gygi1986} and
real-space truncation of the Coulomb interaction~\cite{Spencer2008,Sundararaman2013}.  In
this work, all correlated calculations and their underlying HF calculations are done with
no such corrections.  However, when total energies are required, we separately converge
the HF energy using an exchange interaction with a spherically truncated Coulomb
potential; these calculations are performed with isotropic $k$-point meshes and the
results are extrapolated using a finite-size scaling of the form
$\exp(-aN_k^{1/3})$~\cite{Spencer2008,Sundararaman2013}. 

Finite-size errors in the correlated theories can be analyzed in a similar manner. 
With $k$-point sampling, the MP2 energy per unit cell is approximated by
\begin{equation}
\begin{split}
&E_{\mathrm{MP2}} = \frac{\Omega}{N_k^3} \sum^\prime_{\vk_a\vk_b\vk_i\vk_j}
\sum_{abij} t_{\ii\jj}^{\aa\bb} \\
&\hspace{0em} \times \Bigg[ 2\sum^\prime_{\vG} \rho_{\ii,\aa}(\vG)
        \frac{4\pi}{|\vG-\vq_{ia}|^2} \rho_{\jj,\bb}(\vG_{ijab}-\vG) \\
&\hspace{1em} -\sum^\prime_{\vG} \rho_{\ii,\bb}(\vG)
        \frac{4\pi}{|\vG-\vq_{ib}|^2} \rho_{\jj,\aa}(\vG_{ijba}-\vG) \Bigg]
\end{split}
\end{equation}
where again the term with $G=0$ is neglected when 
$\vq_{ia} = \vk_i-\vk_a = 0$ (first term)
or $\vq_{ib} = \vk_i-\vk_b = 0$ (second term).
Unlike in the case of exchange, these neglected terms vanish at the origin $\vq = 0$
because the orthogonality of orbitals guarantees that
$\rho_{i\vk,a\vk}(\vG=0) = 0$; instead, these terms are associated with an integrand of
$O(q^2)$, leading to an integrated error of $O(N_k^{-1})$.  If the $t$-amplitudes are
correct, then this is the only integration error.  However, the amplitudes 
\begin{equation}
t_{\ii\jj}^{\aa\bb} = \frac{(\ii\aa|\jj\bb)^*}
    {\varepsilon_{\ii}+\varepsilon_{\jj}-\varepsilon_{\aa}-\varepsilon_{\bb}}
\end{equation}
inherit the error of the HF orbital energies, which can be analyzed analogously.
Orthogonality guarantees that unoccupied energies have the same favorable $N_k^{-1}$
error; however, just as for the exchange energy, the occupied orbitals exhibit an
$N_k^{-1/3}$ error~\cite{Aissing1993}.  Without any corrections to the orbital energies,
this pollutes the $t$-amplitudes and thus dominates the error in the correlation energies
(see below).  The inclusion of all four-index integrals in CCSD is similarly responsible
for a $N_k^{-1/3}$ error.

\section{Ground-state results}
\label{sec:results_gs}

We first consider the ground-state properties of diamond and silicon, to establish the
convergence properties of the ground state CC before proceeding to our excited state
studies in the next section. Diamond and silicon share the same crystal structure, with
two atoms per (primitive) unit cell.  Except where otherwise noted, all calculations are
performed with the $T=300$~K experimental lattice constants $a = 3.567$~\AA~and $a =
5.431$~\AA, for diamond and silicon respectively.

We use GTH pseudopotentials~\cite{Goedecker1996,Hartwigsen1998}, explicitly treating four
valence electrons per atom (eight per unit cell), and matching single-particle basis sets,
obtained from the CP2K software package~\cite{VandeVondele2005,VandeVondele2007,CP2K}.  In
the present work, we use pseudopotentials and basis sets that were originally optimized
for use in DFT calculations with the local density approximation (LDA); future work will
consider pseudopotentials and basis sets that are optimized for HF
calculations~\cite{DelBen2012} and we will present a comparison with all-electron
calculations.  The DZV, DZVP, and TZVP basis sets have 8, 13, and 17 orbitals per atom
(twice as many per cell).  The real-space grid spacing used for the integrals was about
0.17~\AA, corresponding to a kinetic energy cutoff of approximately 400 Ry (note that the
same grid is used to resolve the orbitals, Eq.~(\ref{eq:ao_pw}), and the pair densities,
Eq.~(\ref{eq:density_pw})).  All calculations are converged, with respect to this grid, to
an accuracy better than $10^{-4}$ au per atom.  In all ground-state calculations, the
Brillouin zone was sampled from a uniform $\Gamma$-centered grid.  All calculations were
performed with the PySCF software package~\cite{PySCF}.


\begin{figure}[b]
\centering
\includegraphics[scale=1.0]{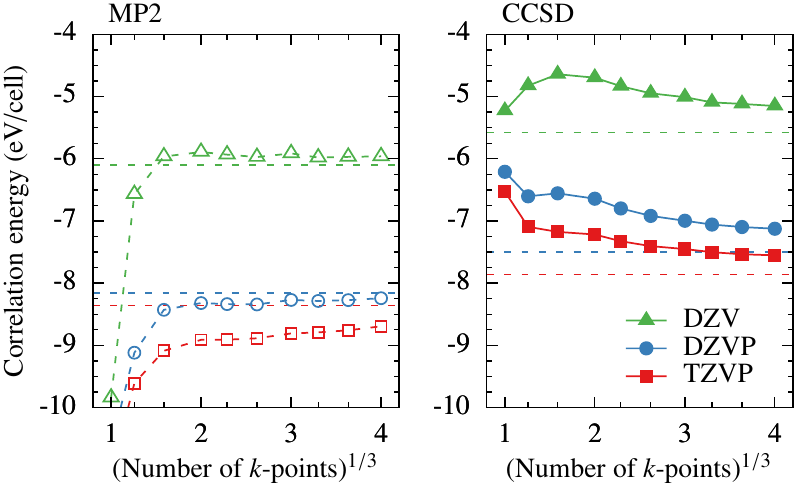}
\caption{
Correlation energy per unit cell of diamond calculated with MP2 (left, open symbols) and
CCSD (right, filled symbols). Dashed lines indicate the $N_k\rightarrow \infty$
extrapolated values assuming a finite-size scaling of the form $N_k^{-1/3}$.
}
\label{fig:gs_energy}
\end{figure}

As described above, total convergence in periodic solids must be achieved with respect to
the single-particle basis and the sampling of the Brillouin zone (finite-size effects).
In Fig.~\ref{fig:gs_energy} we show the convergence of the ground-state correlation energy
of diamond, at the MP2 and CCSD levels of theory, as a function of the size of the
single-particle basis and the number of $k$-points sampled in the Brillouin zone.  As
expected from molecular calculations with comparable basis sets, the total correlation
energies are not converged with respect to the single-particle basis; the correlation
energy difference between results obtained with the DZVP and TZVP basis sets is
0.4--0.5~eV per unit cell (half that per atom), to be compared to the HF energy
differences, which are less than 0.1~eV (not shown). In this work, we do not pursue a
complete basis set limit extrapolation based only on double-zeta and triple-zeta basis
sets. It can be expected, however, that this basis set error is local in character and can
be corrected using standard explicit $r_{12}$ correlation techniques~\cite{Klopper2006}.
Furthermore, we observe that the basis set corrections to the correlation energy are
similar for MP2 and CCSD (as in molecules), and we could thus similarly use complete basis
set limit MP2 calculations to correct the CCSD results.

With regards to finite-size effects, we observe that in small basis sets the MP2
correlation energy converges more quickly than the CCSD energy with the number of
$k$-points. However, for the largest (TZVP) basis set, the energy convergence is similar.
In Tab.~\ref{tab:gs_energy}, we give the HF and correlation energies obtained with a
$3\times 3\times 3$ and $4\times 4\times 4$ sampling of the Brillouin zone; the largest CC
calculation presented here, using a $4\times 4\times 4$ sampling of the Brillouin zone and
the TZVP basis, constitutes a canonical CCSD calculation of 256 electrons in 2,176
orbitals, demonstrating the savings provided by incorporating periodic symmetry.

\begin{table}[t]
\centering
\begin{tabular*}{0.48\textwidth}{@{\extracolsep{\fill}} llccc }
\hline\hline
&                   & DZV    & DZVP     & TZVP        \\
\hline
HF & & & & \\
& Extrap, $\exp(-aN_k^{1/3})$ & -301.05 & -301.86 &  -301.95 \\
&&&&\\
MP2 & & & &\\
& $3\times 3\times 3$   & -5.91 & -8.27 & -8.81 \\
& $4\times 4\times 4$   & -5.96 & -8.24 & -8.70 \\
& Extrap, $N_k^{-1/3}$ & -6.10 & -8.16 & -8.36 \\
&&&&\\
CCSD & & & &\\                                 
& $3\times 3\times 3$   & -5.01 & -7.00 & -7.45 \\
& $4\times 4\times 4$   & -5.15 & -7.12 & -7.55 \\
& Extrap, $N_k^{-1/3}$ & -5.58 & -7.50 & -7.86 \\
\hline\hline
\end{tabular*}
\caption{
Hartree-Fock total energy per unit cell and MP2/CCSD correlation energy per unit cell of
diamond in eV.  
}
\label{tab:gs_energy}
\end{table}

\begin{figure}[b]
\centering
\includegraphics[scale=1.0]{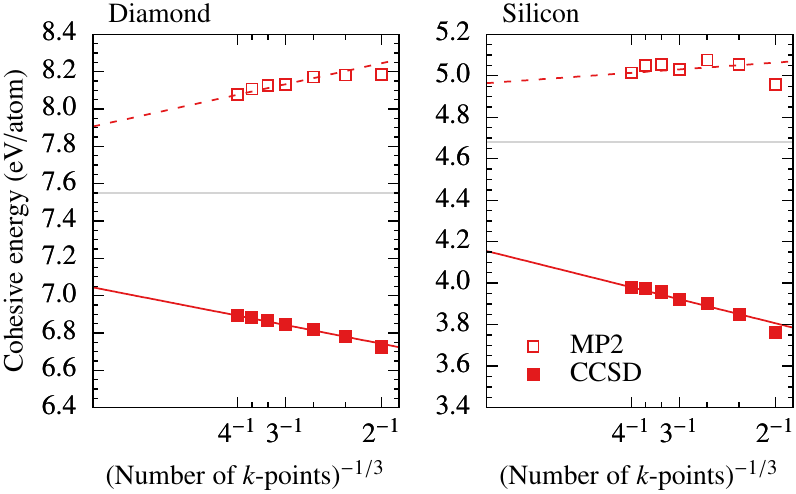}
\caption{
Cohesive energy of diamond (left) and silicon (right), calculated with the TZVP basis. For
each $k$-point mesh, MP2 and CCSD correlation energies are added to a separately converged
HF crystal energy.  The correlation component of the cohesive energy is extrapolated
assuming a finite-size error that scales like $N_k^{-1/3}$.  Zero-point corrected
experimental results~\cite{Schimka2011} are shown as a thin grey line.  Predicted and
experimental values are given in Tab.~\ref{tab:properties}.
}
\label{fig:cohesive}
\end{figure}

In Tab.~\ref{tab:gs_energy}, we also report the thermodynamic limit based on an
extrapolation of the form $N_k^{-1/3}$, as discussed in Sec.~\ref{sec:finite}.  This
extrapolation is graphically demonstrated in Fig.~\ref{fig:cohesive}, which shows the
cohesive energy (the energy-per-atom difference between the crystal and isolated atom) of
diamond and silicon as a function of the number of $k$-points sampled using the TZVP
basis.  Specifically, the fitting is restricted to the two largest isotropic $k$-point
meshes, $n\times n\times n$ with $n=3,4$.  Atomic energies of open-shell C and Si were
calculated with unrestricted MP2 and CCSD in the field of crystalline basis functions (to
account for basis set superposition error) and the HF contribution to the crystal energy
is separately converged with respect to the number of $k$-points, in the manner described
in Sec.~\ref{sec:finite}.  Predicted cohesive energies are given in
Tab.~\ref{tab:properties}.

\begin{table}[t]
\centering
\begin{tabular*}{0.48\textwidth}{@{\extracolsep{\fill}} llcccccc }
\hline\hline
&           & \multicolumn{2}{c}{$a$ (\AA)} & \multicolumn{2}{c}{$B$ (GPa)} & \multicolumn{2}{c}{$\Delta E$ (eV/atom)} \\
\hline
C &     &            &         &     &       &      &         \\
& HF    &      3.527 & (3.552) & 507 & (495) & 5.36 & (5.28)  \\
& MP2   &      3.545 & (3.553) & 436 & (450) & 7.91 & (7.97,8.039) \\
& CCSD  &      3.539 & ---     & 463 & ---   & 7.04 & (7.295) \\
& Experiment & \multicolumn{2}{c}{3.553} & \multicolumn{2}{c}{455} & \multicolumn{2}{c}{7.55} \\
&&&&&&&\\
Si &    &            &         &     &       &      &        \\
& HF    &      5.435 & (5.512) & 107 & (103) & 3.03 & (2.97) \\
& MP2   &      5.347 & (5.415) & 101 & (100) & 4.96 & (5.05) \\
& CCSD  &      5.393 & ---     & 103 & ---   & 4.15 & ---    \\
& Experiment & \multicolumn{2}{c}{5.421} & \multicolumn{2}{c}{101} & \multicolumn{2}{c}{4.68} \\
\hline\hline
\end{tabular*}
\caption{
Lattice constant $a$, bulk modulus $B$, and cohesive energy $\Delta E$ of diamond and
silicon using the TZVP single-particle basis.  Lattice constant and bulk modulus were
obtained from fits to calculations with a $3 \times 3 \times 3$ $k$-point mesh.  Crystal
energy contributions to the cohesive energy are calculated at the $T=300$~K lattice
constant and extrapolated to $N_k \rightarrow \infty$.  Values given in parentheses are
those calculated with the plane-wave PAW approach as reported in
Refs.~\onlinecite{Gruneis2010,Booth2013}. All experimental values have been corrected for
zero-point vibrational effects~\cite{Schimka2011}.
}
\label{tab:properties}
\end{table}

Our cohesive energies for diamond, 7.91 eV/atom (MP2) and 7.04 eV/atom (CCSD), are in
reasonable agreement with those of previous periodic MP2 calculations, 7.97
eV/atom~\cite{Gruneis2010} and 8.039 eV/atom~\cite{Booth2013}, and of a previous CCSD
calculation, 7.295 eV/atom~\cite{Booth2013}.  Similarly, our MP2 cohesive energy for
silicon, 4.96 eV/atom, is in good agreement with the previously reported MP2 value, 5.05
eV/atom~\cite{Gruneis2010}.  The discrepancies in our data of order 0.1--0.2~eV/atom
likely originate from a combination of the finite basis set and treatment of core
electrons.  The calculations reported in Refs.~\onlinecite{Gruneis2010,Booth2013} employ a
plane-wave basis set and the projector-augmented wave (PAW) method, which differs from the
norm-conserving pseudopotentials used here.  Furthermore, we always find that the absolute
value of the correlation energy given by MP2 is larger than that given by CCSD, which is
the opposite of that observed in Ref.~\onlinecite{Booth2013}; this difference may also be
attributable to the difference between the PAW and pseudopotential approximations.  

\begin{figure}[t]
\centering
\includegraphics[scale=1.0]{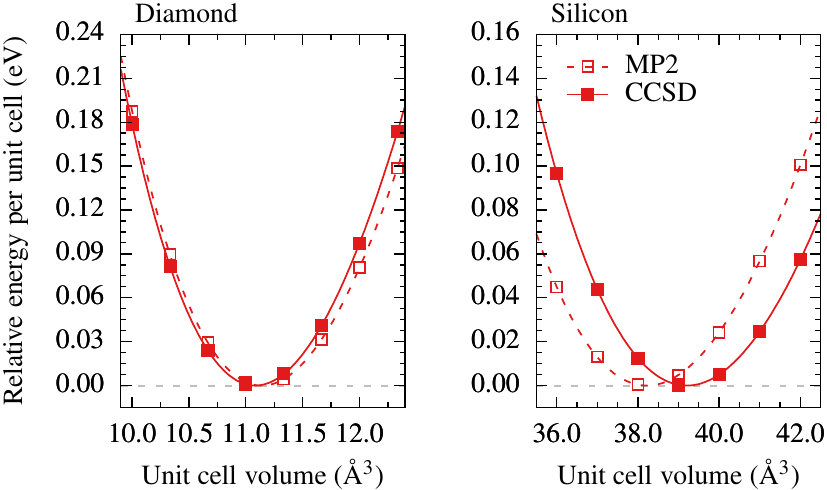}
\caption{
Equation of state of diamond (left) and silicon (right) with the TZVP basis and a $3\times
3\times 3$ sampling of the Brillouin zone.
}
\label{fig:eos}
\end{figure}

Using a $3\times 3\times 3$ $k$-point mesh and the TZVP basis, we have further calculated
the equation of state of diamond and silicon, shown in Fig.~\ref{fig:eos}.  We performed
at least seven ground-state calculations with unit cell volumes varying by about $\pm
10\%$ from the equilibrium value and fit the results to a third-order Birch-Murnaghan
form, leading to predicted (zero-temperature) lattice constants and bulk moduli, given in
Tab.~\ref{tab:properties}.  Again, the HF and MP2 structural predictions are in good
agreement with previous values~\cite{Gruneis2010}, and the observed discrepancy at the HF
level provides some measure of the error incurred due to the use of pseudopotentials.
Ultimately, the overall good agreement between our results and those of
Refs.~\onlinecite{Gruneis2010,Booth2013} suggests that the correlation energy differences
required for structural properties and cohesive energies are reasonably converged in both
implementations, with respect to both the basis and $k$-point sampling.  At least for the
two semiconductors studied here, MP2 and CCSD provide a similar level of accuracy when
compared to experimental values determined by ground-state energetics.

\section{Excited-state results}
\label{sec:results_eom}

\begin{figure}[b]
\centering
\includegraphics[scale=1.0]{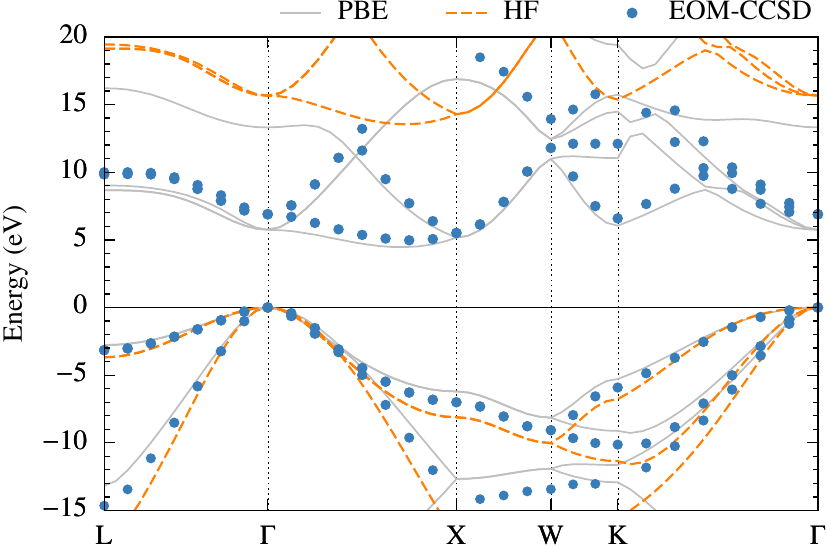}
\caption{
Band structure of diamond calculated with DFT (PBE), HF, and EOM-CCSD, using the DZVP
single-particle basis and a $3\times 3\times 3$ $k$-point mesh.  
}
\label{fig:bands_c}
\end{figure}

\begin{figure}[b]
\centering
\includegraphics[scale=1.0]{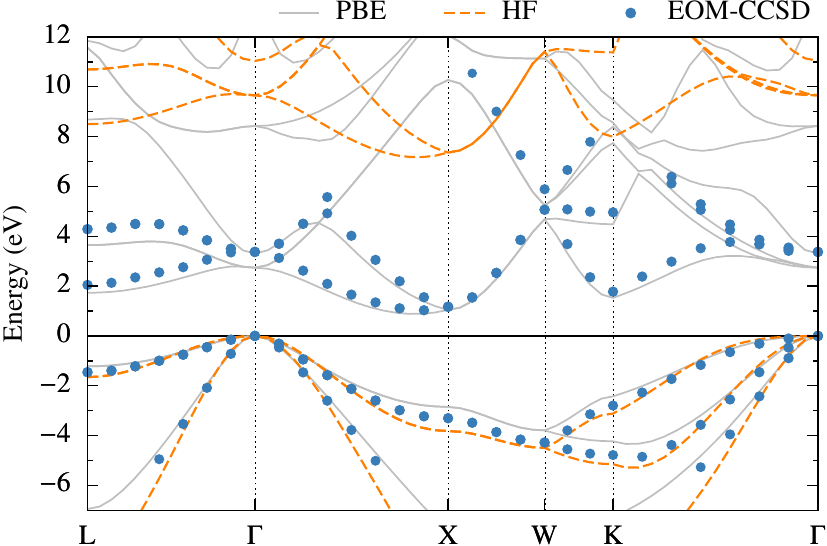}
\caption{
The same as in Fig.~\ref{fig:bands_c}, but for silicon.
}
\label{fig:bands_si}
\end{figure}

We now turn to the excited-state and spectral calculations using the EOM-CCSD formalism.
As discussed in Sec.~\ref{sec:cc}, we compute the quasiparticle excitations in the
ionization potential (IP) and electron affinity (EA) charged sectors.  To obtain
continuous bands in conventional band structure calculations with HF or DFT, an initial
self-consistent calculation is performed with a fixed (and potentially coarse) $k$-point
grid, which provides an approximate representation of the density matrix or density; see
Eq.~(\ref{eq:dm}). From this latter object, the one-body Fock-like matrix can be
constructed at \textit{arbitrary} $k$-points and diagonalized, leading to
non-self-consistent eigenvalues.  To construct a continuous band-structure in EOM-CCSD, we
here perform a large series of (independent) calculations with a $k$-point mesh that is
shifted to include the desired $k$-point.  Although this means that certain calculations
along the band path will be performed with a sub-optimal mesh, the results will converge
properly in the limit of dense $k$-point sampling.  We only present results here for a
band structure path through the Brillouin zone, but the same technique could be applied to
calculate the full quasiparticle density of states.

The band structures of diamond and silicon are shown in Figs.~\ref{fig:bands_c} and
\ref{fig:bands_si} using DFT with the PBE functional~\cite{Perdew1996}, HF, and EOM-CCSD
using the DZVP basis and a $3\times 3\times 3$ $k$-point mesh.  Qualitatively, the
EOM-CCSD result reproduces the expected behavior: it predicts a band gap that is slightly
larger than that of PBE and significantly smaller than that of HF.  Similarly, the
EOM-CCSD band widths are intermediate between the two mean-field results.

\begin{figure}[t]
\centering
\includegraphics[scale=1.0]{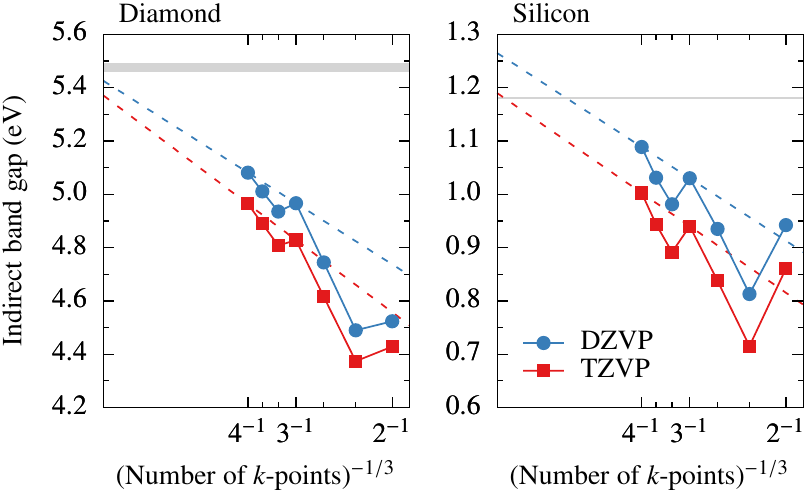}
\caption{
Indirect band gap of diamond (left) and silicon (right) calculated with EOM-CCSD.  Dashed
lines indicate extrapolation of the form $N_k^{-1/3}$ based on the isotropic $3\times
3\times 3$ and $4\times 4\times 4$ sampling meshes; grey horizontal lines indicate
experimental values.  For the TZVP data, the extrapolation predicts for diamond a band gap
of 5.37 eV compared to the experimental value of 5.45--5.50
eV~\cite{Clark1964,ODonnell1991}, and for silicon a band gap of 1.19~eV, compared to the
experimental value of 1.17 eV~\cite{Bludau1974,ODonnell1991}.
}
\label{fig:bandgap}
\end{figure}

To study the convergence of the excited-state structure, we present in
Fig.~\ref{fig:bandgap} the calculated value of the indirect (minimum) band gap in diamond
and silicon, as a function of the basis set and number of $k$-points sampled.  Since we do
not calculate the full band path due to its high cost, the indirect band gap is evaluated
from two independent calculations: an IP-EOM-CCSD calculation with a $\Gamma$-centered
mesh and an EA-EOM-CCSD calculation with a $k$-point mesh centered at $\sim 85$\% of the
$\Gamma-X$ line (near the zone boundary).  We see that larger single-particle basis sets
favor a smaller bandgap and finer $k$-point meshes favor a larger band gap.  The band gap
difference between calculations using the DZVP and TZVP basis sets is about 0.1~eV, and
relatively independent of the number of $k$-points sampled.

The band gap convergence with the number of $k$-points is slow and oscillatory (when
including anisotropic $k$-point meshes), a behavior inherited from the underlying HF band
structure (not shown). In particular, we retain the slow convergence arising from the
$\vG=0,\ \vq\to 0$ contributions to the energy discussed in Sec.~\ref{sec:hf}, which
appears to lead to an $N_k^{-1/3}$ scaling of the error in the bandgap. Similarly to the
ground-state energy however, the finite-size effects are much smaller in the correlated
band gap than in the HF band gap: in diamond, the HF band gap changes by about 7~eV during
convergence, while the EOM-CCSD band gap changes by less than 1~eV.

Like in our treatment of the ground-state and cohesive energy, we have applied a
$N_k^{-1/3}$ finite-size scaling to the EOM-CCSD band gaps, restricted to the two largest
isotropic $k$-point meshes.  Within the approximations of our approach (GTH
pseudopotentials, finite basis, and EOM-CCSD correlation) we predict band gaps of 5.37~eV
and 1.19~eV, for diamond and silicon respectively; the experimental $T=0$ band gaps are
5.45--5.50 eV~\cite{Clark1964,ODonnell1991} and 1.17 eV~\cite{Bludau1974,ODonnell1991},
which are corrected for excitonic and finite-temperature effects, but not for the more
challenging electron-phonon (vibrational) effects.  For diamond in particular, a
zero-point band gap renormalization of 0.4--0.6~eV has been
estimated~\cite{Cardona2005,Giustino2010,Antonius2014}, which would suggest an
experimental electronic band gap of 5.8~eV or more.  For comparison, tightly converged
all-electron $G_0W_0$ calculations predict band gaps of 5.61--5.62~eV and
1.10--1.11~eV~\cite{Friedrich2010,Klimes2014}, which are quite close to $G_0W_0$
calculations with norm-conserving pseudopotentials~\cite{Chen2015}, but 0.1--0.2~eV
smaller than all-electron self-consistent $GW$ calculations~\cite{Shishkin2007PRB} and
0.2--0.3 eV smaller than vertex-corrected calculations~\cite{Gruneis2014}.  Interestingly,
although MP2 predicts ground-state structural properties that are on par with CCSD, it
massively underestimates the band gaps of these two materials, giving 1.9~eV for diamond
and $-1.2$~eV for silicon~\cite{Gruneis2010}, which was attributed to the large
polarizability of these materials.

\section{Discussion and conclusion}
\label{sec:conc}
\raggedbottom

In this work, we have presented the first-ever results from canonical Gaussian-based
coupled-cluster theory for three-dimensional solids.  We discussed concomitant finite-size
errors in the mean-field and correlated calculations.  Based on ground-state
coupled-cluster calculations, we presented the lattice constant, bulk modulus, and
cohesive energies of diamond and silicon; these results are in good agreement with limited
existing data based on the plane-wave PAW approach~\cite{Gruneis2010,Booth2013}.  We
further presented the first results of excited-state (equation-of-motion) coupled-cluster
theory for the band structure and band gap of the same two semiconductors.  Our predicted
minimum band gaps, 5.37~eV for diamond and 1.19~eV for silicon, are in good agreement with
experimental values.

A great deal of work remains to be done to make coupled-cluster calculations on
weakly-correlated solids  routine.  In future work, we will carefully compare the results
of pseudopotential and all-electron calculations, which represents an unquantified source
of error.  Parallel efforts are aimed at reducing the cost and increasing our
understanding of these calculations, for example through local correlation approaches,
explicitly correlated formulations, perturbative corrections due to triple excitations,
and applications to a wider range of insulating and metallic materials.

\acknowledgments

J.M., Q.S., and G.K.-L.C.~were supported by the US Department of Energy, Office of Science
under Award Number DE-SC0008624. Additional support was provided by the Simons Foundation,
via the Collaboration on the Many-Electron Problem, and a Simons Investigatorship in
Theoretical Physics.  T.C.B. was supported by the Princeton Center for Theoretical Science
and by start-up funds from the University of Chicago.

\pagebreak
\begin{widetext}

\appendix*

\section{IP/EA-EOM-CCSD equations for periodic systems}

Assuming a closed-shell reference, the spatial-orbital IP-EOM-CCSD amplitude equations are
given by~\cite{Nooijen1993ccgf}
\begin{subequations}
\begin{align}
\begin{split}
(\bar{H}R^-)_{\ii} &= \sum_{k}\sum_{\vk_k}^\prime -U_{\kk \ii} r_{\kk}
    + \sum_{ld} \sum_{\vk_l \vk_d}^\prime U_{\ll \dd}
        (2 \rip{\ii}{\ll}{\dd} - \rip{\ll}{\ii}{\dd}) \\
&\hspace{1em} - \sum_{kld} \sum_{\vk_k \vk_l \vk_d}^\prime
    (2 W_{\kk \ll \ii \dd} - W_{\ll \kk \ii \dd} ) \rip{\kk}{\ll}{\dd}
\end{split}\\
\begin{split}
\label{eq:r2_ip}
(\bar{H }R^-)_{\ii \jj}^{\hphantom{\ii}\bb} &=
    \sum_{d} \sum_{\vk_d}^\prime U_{\bb \dd } \rip{\ii}{\jj}{\dd}
    - \sum_{l} \sum_{\vk_l}^\prime U_{\ll \ii} \rip{\ll}{\jj}{\bb}
    - \sum_{l} \sum_{\vk_l}^\prime U_{\ll\jj} \rip{\ii}{\ll}{\bb} \\
&\hspace{1em} - \sum_{k} \sum_{\vk_k}^\prime W_{\kk \bb \ii \jj} r_{\kk}
    + \sum_{ld} \sum_{\vk_l \vk_d}^\prime (2 W_{\ll \bb \dd \jj} - W_{\bb \ll \dd \jj} )
                    \rip{\ii}{\ll}{\dd}
    - \sum_{ld} \sum_{\vk_l \vk_d}^\prime W_{\ll \bb \dd \jj} \rip{\ll}{\ii}{\dd}\\
&\hspace{1em} + \sum_{kl} \sum_{\vk_k \vk_l}^\prime W_{\kk \ll \ii \jj} \rip{\kk}{\ll}{\bb}
    -\sum_{kd} \sum_{\vk_k \vk_d}^\prime W_{\kk \bb \ii \dd} \rip{\kk}{\jj}{\dd} \\
&\hspace{1em} - \sum_{c}\sum_{\vk_c}^\prime \left[
        \sum_{kld} \sum_{\vk_k \vk_l \vk_d}^\prime (2 W_{\ll \kk \dd \cc} - W_{\kk \ll \dd \cc})
            \rip{\kk}{\ll}{\dd} \right] t_{\ii \jj}^{\cc \bb},
\end{split}
\end{align}
\end{subequations}
and the EA-EOM-CCSD equations are given by~\cite{Nooijen1995}
\begin{subequations}
\begin{align}
\begin{split}
(\bar{H}R^+)^{\aa} &= \sum_{c} \sum_{\vk_c}^\prime U_{\aa \cc} r^{\cc}
    + \sum_{ld} \sum_{\vk_l \vk_d}^\prime U_{\ll \dd}
        (2 \rea{\ll}{\aa}{\dd} - \rea{\ll}{\dd}{\aa}) \\
&\hspace{1em} + \sum_{lcd} \sum_{\vk_l \vk_c \vk_d}^\prime
    (2 W_{\aa \ll \cc \dd} - W_{\aa \ll \dd \cc}) \rea{\ll}{\cc}{\dd}
\end{split}\\
\begin{split}
\label{eq:r2_ea}
(\bar{H}R^+)_{\hphantom{\aa}\jj}^{\aa \bb} &=
    - \sum_{l} \sum_{\vk_l}^\prime U_{\ll \jj} \rea{\ll}{\aa}{\bb}
    + \sum_{c} \sum_{\vk_c}^\prime U_{\aa \cc} \rea{\jj}{\cc}{\bb}
    + \sum_{d} \sum_{\vk_d}^\prime U_{\bb \dd} \rea{\jj}{\aa}{\dd} \\
&\hspace{1em} + \sum_{c} \sum_{\vk_c}^\prime W_{\aa \bb \cc \jj} r^{\cc}
    + \sum_{ld} \sum_{\vk_l \vk_d}^\prime (2 W_{\ll \bb \dd \jj} - W_{\bb \ll \dd \jj})
                    \rea{\ll}{\aa}{\dd}
    - \sum_{lc} \sum_{\vk_l \vk_c}^\prime W_{\aa \ll \cc \jj} \rea{\ll}{\cc}{\bb} \\
&\hspace{1em} - \sum_{lc} \sum_{\vk_l \vk_c}^\prime W_{\bb \ll \jj \cc} \rea{\ll}{\cc}{\aa}
    + \sum_{cd} \sum_{\vk_c \vk_d}^\prime W_{\aa \bb \cc \dd} \rea{\jj}{\cc}{\dd} \\
&\hspace{1em} - \sum_{k} \sum_{\vk_k}^\prime \left[
        \sum_{lcd} \sum_{\vk_l \vk_c \vk_d}^\prime (2 W_{\kk \ll \cc \dd} - W_{\kk \ll \dd \cc})
            \rea{\ll}{\cc}{\dd}\right] t_{\kk \jj}^{\aa \bb}.
\end{split}
\end{align}
\end{subequations}
As described in the text, primed summations indicate that one of the listed momenta is
fixed by momentum conservation.  The intermediates $U$ and $W$ are the usual ones arising
in coupled-cluster theory from the effective Hamiltonian $\bar{H}$~\cite{Nooijen1993ccev},
and their efficient calculation also accounts for momentum conservation.

\end{widetext}

%

\end{document}